\documentclass[sigconf,10pt]{acmart}
\usepackage{color}
\definecolor{RED}{rgb}{1,0,0}
\definecolor{BLUE}{rgb}{0,0,1}
\definecolor{PURPLE}{rgb}{1,0,1}
\usepackage{enumitem}
\newcommand{\jiangkai}[1]{{#1}}
\newcommand{\jk}[1]{{#1}}
\newcommand{\wu}[1]{{#1}}
\newcommand{\wuu}[1]{{#1}}
\title{Chat with AI: The Surprising Turn of Real-time Video Communication from Human to AI}

\author{{Jiangkai Wu}, 
{Zhiyuan Ren}, 
{Liming Liu},
{Xinggong Zhang}}
\affiliation{
  {Peking University}%
  \country{}%  
}
\renewcommand{\shortauthors}{Jiangkai Wu et al.}

\begin{abstract}
AI Video Chat emerges as a new paradigm for Real-time Communication (RTC), where one peer is not a human, but a Multimodal Large Language Model (MLLM). This makes interaction between humans and AI more intuitive, as if chatting face-to-face with a real person. However, this poses significant challenges to latency, because the MLLM inference takes up most of the response time, leaving very little time for video streaming. \textit{Due to network uncertainty, transmission latency becomes a critical bottleneck preventing AI from being like a real person.} To address this, we call for AI-oriented RTC research, \textit{exploring the network requirement shift from "humans watching video" to "AI understanding video".} We begin by recognizing the main differences between AI Video Chat and traditional RTC. Then, through prototype measurements, we identify that ultra-low bitrate is a key factor for low latency. To reduce bitrate dramatically while maintaining MLLM accuracy, we propose \textit{Context-Aware Video Streaming} that recognizes the importance of each video region for chat and allocates bitrate almost exclusively to chat-important regions. To evaluate the impact of video streaming quality on MLLM accuracy, we build the first benchmark, named \textit{\textbf{De}graded \textbf{Vi}deo Understanding \textbf{Bench}mark (\textbf{DeViBench})}.
Finally, we discuss some open questions and ongoing solutions for AI Video Chat. \jiangkai{DeViBench is open-sourced at: \href{https://github.com/pku-netvideo/DeViBench}{\textcolor{magenta}{https://github.com/pku-netvideo/DeViBench}}.}
\end{abstract}

\copyrightyear{2025}
\acmYear{2025}
\setcopyright{acmlicensed}\acmConference[HotNets '25]{The 24th ACM
Workshop on Hot Topics in Networks}{November 17--18, 2025}{College Park,
MD, USA}
\acmBooktitle{The 24th ACM Workshop on Hot Topics in Networks (HotNets
'25), November 17--18, 2025, College Park, MD, USA}
\acmDOI{10.1145/3772356.3772390}
\acmISBN{979-8-4007-2280-6/2025/11}

\begin{document}



\maketitle

\vspace{-4mm}
\begin{figure}[h]
\centering
\setlength{\abovecaptionskip}{0mm}
\includegraphics[width=0.8\linewidth]{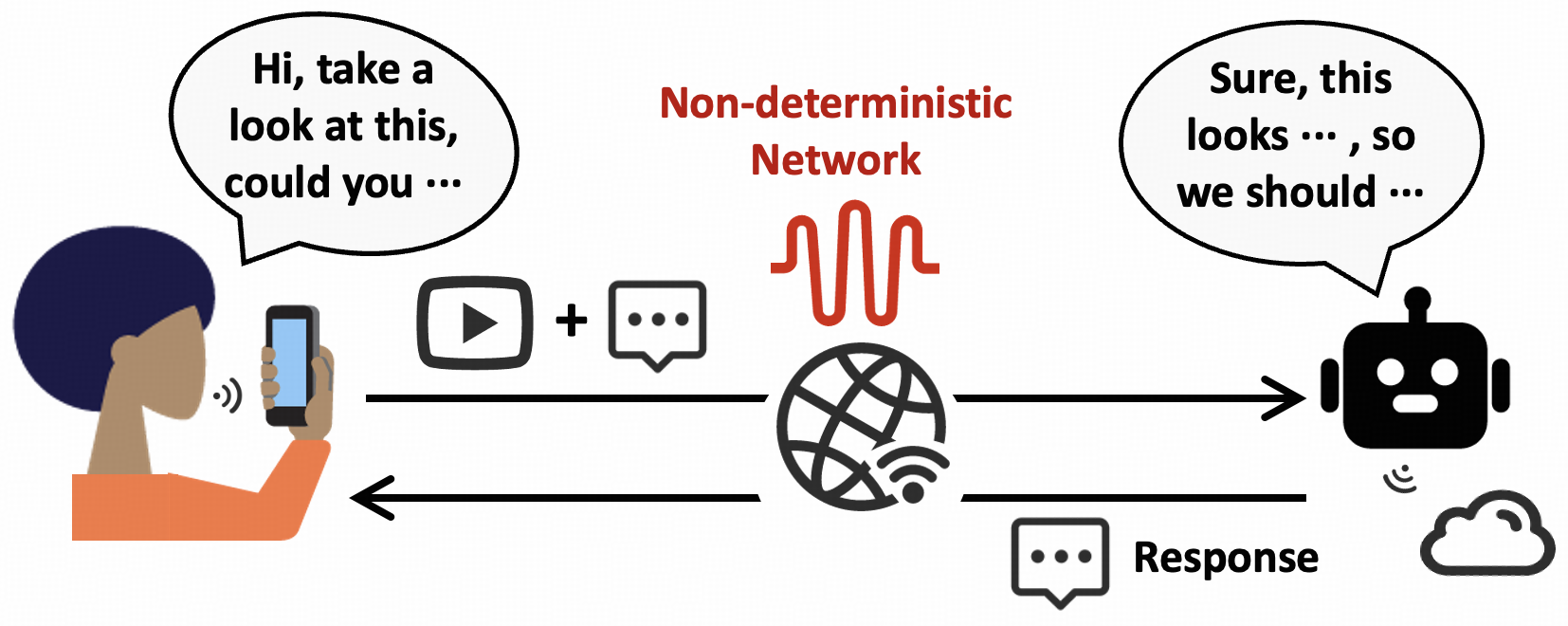}
\caption{AI Video Chat is a new paradigm for real-time communication. The user sends video and audio to the AI for thinking. The AI feeds back to the user. Low latency is crucial for making AI act like a real person.}
\label{fig:teaser}
% \vspace{-4mm}
\end{figure}

\vspace{-4mm}
\section{Introduction}

AI Video Chat is a new paradigm for Real-time Communication (RTC). Since OpenAI released GPT-4o~\cite{hurst2024gpt}, Multimodal Large Language Models (MLLMs) have been continuously emerging, such as Qwen2.5-Omni~\cite{xu2025qwen2}, Gemini-1.5~\cite{team2024gemini}, VITA-1.5~\cite{fu2025vita}, and OmniLive~\cite{zhang2024internlm}. Compared to traditional LLMs, MLLMs enable users to directly input video and audio for interaction, rather than just text. This makes the interaction between humans and AI more intuitive as if chatting face-to-face with a real person~\cite{chen2024videollm}. However, MLLMs require high-performance computing devices to support real-time inference (such as 8*A100 GPUs~\cite{qiu2025modservemodalitystageawareresource}). Mobile devices (like phones or smart glasses) cannot meet the computing requirements, which makes MLLMs inevitably deployed in the cloud. So in existing systems, the client sends user video and audio to the cloud for MLLM inference, and the cloud then feeds back responses to users, as shown in Figure~\ref{fig:teaser}.

AI Video Chat raises significant challenges to RTC transmission latency. To ensure a fluent interactive experience, the response latency of video chat needs to remain below 300 ms~\cite{lai2022spacertc}. In traditional video chat, the human peer on the other side can respond instantly. Therefore, \jiangkai{response latency largely stems from RTC's end-to-end latency (including capture, transmission, decoding, and playback buffer latencies). Among them, transmission latency is sensitive to network conditions and continuously increases when the network deteriorates.} To reduce transmission latency, current state-of-the-art RTC frameworks (such as WebRTC~\cite{carlucci2016analysis}) adopt technologies like Adaptive Bitrate (ABR)~\cite{mao2017neural,yan2020learning,akhtar2018oboe,huang2014buffer,jiang2012improving}, Congestion Control~\cite{cardwell2017bbr,dong2015pcc,dong2018pcc,meng2024feedback}, and Forward Error Correction (FEC)~\cite{meng2024hairpin,an2025tooth,rudow2023tambur,cheng2024grace,li2023reparo} to satisfy user experience. However, in AI Video Chat, responses are generated through MLLM in an autoregressive manner, which is time-consuming. Even when inputting only audio tokens, the computational latency is at least 232 ms~\cite{hurst2024gpt}. To constrain the response latency below 300 ms, \jiangkai{even ignoring other latencies in the RTC pipeline,} the time left for transmission is at most 68 ms, which is difficult to guarantee. So the large latency makes users clearly feel that the other side is not a real person.

\textbf{Is it possible to reduce the latency of AI Video Chat to an extremely low level? 
This vision allows us to grasp the "holy grail"~\cite{chen2024videollm} of AI research from the perspective of network systems: making AI like real humans.}

To realize this vision, we call for \jiangkai{AI-oriented RTC research}, exploring the network requirement shift from "humans watching video" to "AI understanding video". We begin by recognizing the main differences between AI Video
Chat and traditional RTC: \textit{First, QoE changes from human perceptual quality to MLLM response accuracy.} \textit{Second, jitter has no impact.} \textit{Third, the receiver throughput is far lower than the sender throughput.} \textit{Fourth, uplink is more pressing than Downlink.} Then, through prototype measurements, we identify a key factor for low latency: \textit{ultra-low bitrate}. Based on two insights, we make contributions:
% \vspace{-6mm}
\begin{itemize}[left=0pt]
\item \textbf{Video should be Context-Aware (\S\ref{sec:method_1}).} To reduce bitrate dramatically while maintaining MLLM accuracy, we propose \textit{Context-Aware Video Streaming}, allocating more bitrate to chat-important video regions while allocating as little bitrate as possible
to chat-irrelevant regions.
% \item \textbf{Frame Rate is FEC (\S\ref{sec:method_2}).} Since MLLMs process videos at a very low frame rate, most received frames are redundant. Although redundant frames cause bitrate waste, they can also substitute for lost/delayed frames. So we propose \textit{Loss-Resilient Adaptive Frame Rate} to simultaneously minimize bitrate waste and packet retransmission.
% \vspace{4mm}
\item \textbf{The First Benchmark (\S\ref{sec:method_3}).} Considering that there is no benchmark that can evaluate the impact of video quality on MLLM accuracy, we propose the first one, named \textit{\textbf{De}graded \textbf{Vi}deo Understanding \textbf{Bench}mark} (\textit{\textbf{DeViBench}}).
\end{itemize}

% \vspace{-4mm}
\vspace{-3mm}
\section{Motivation}

\subsection{Main differences between AI Video Chat and traditional RTC}
\label{sec:moti_1}

\noindent \textbf{QoE changes from human perceptual quality to MLLM response accuracy.} In traditional RTC, QoE focuses on measuring the perceptual quality of human eyes. For example, using penalty terms like stalling time~\cite{cheng2024grace,yan2020learning} or quality variance~\cite{chen2024soda} to measure temporal stability. Using SSIM~\cite{yan2020learning} or VMAF~\cite{vmaf} to measure visual quality. But in AI Video Chat, the viewer of the video changes from humans to MLLMs. At this point, most perception-related metrics are no longer needed, and the optimization objective of QoE becomes the accuracy and latency of MLLM's responses. So the RTC strategy can undergo a significant turn. For example, to reduce latency, temporal stability (such as frequent bitrate adjustments) and visual quality (like lowering bitrate in Figure~\ref{fig:bitrate_reduce_accuracy}) can be sacrificed, as long as accuracy is enough.

\noindent \jiangkai{\textbf{Jitter has no impact.} Due to network uncertainties (congestion, packet loss, etc.), even when the sender transmits frames at fixed time intervals, the time intervals of received frames will experience jitter. Direct playback will result in uneven video speed, causing stuttering. Therefore, traditional RTC employs a jitter buffer~\cite{zhao2024jitbright}, trading latency for smoothness. For MLLMs, the difference is that their perception of time does not rely on real physical time, but rather on positional encoding computation~\cite{xu2025qwen2,fu2025vita,zhang2024internlm}. Positional encoding is only associated with the frame's capture timestamp and unrelated to the actual receiving time; thus, jitter has no impact on the MLLM's perception of the video. It means that in AI video chat, the buffer can be removed to reduce the latency.}

\begin{figure}
\setlength{\abovecaptionskip}{1.mm}
    \centerline{\includegraphics[width=0.66\linewidth]{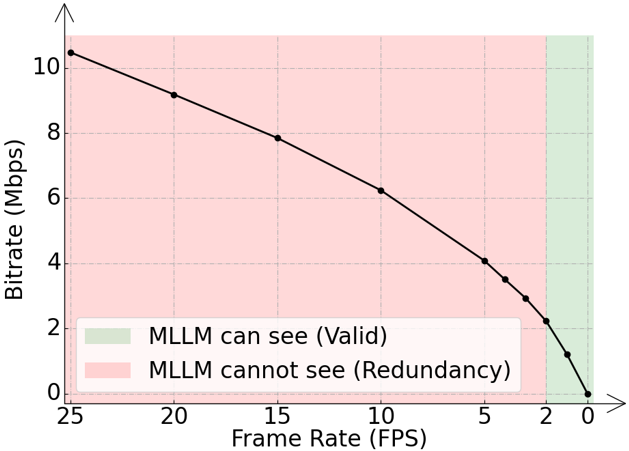}}
    \caption{MLLM processes video at a very low frame rate (green), so most frames are redundancy (red).}
    \vspace{-4mm}
    \label{fig:fps}
\end{figure}

\noindent \textbf{The receiver throughput is far lower than the sender throughput.} In traditional RTC, the data throughput at the receiver is comparable to that at the sender, for example, both are 1920*1080 resolution at a 30 FPS frame rate. However, in AI Video Chat, MLLM is limited by context length (finite number of tokens) and real-time inference, and cannot fully process the received video. Therefore, the received video needs to be actively downsampled before being fed to the MLLM. In terms of frame rate, existing AI Video Chat systems support a maximum processing rate of only 2 FPS~\cite{xu2025qwen2,wang2025omnimmi,fu2025vita}. In terms of resolution, regardless of how high the original resolution is, it will be downsampled to no more than 602,112 pixels~\cite{xu2025qwen2}. So traditional RTC contains massive redundancy that MLLMs cannot perceive, as shown in Figure~\ref{fig:fps}.

\noindent \textbf{Uplink is more pressing than Downlink.} In traditional RTC, each peer is both video sender and receiver. In contrast, AI Video Chat is unidirectional video transmission, where the user only acts as the video sender and MLLM only acts as the video receiver. MLLM sends responses to the user in the form of audio or text, and these representations have much lower bitrates than video. Thus, uplink needs better network conditions than downlink, for example, larger bandwidth.

\vspace{-0mm}
\subsection{What factors affect the transmission latency of AI Video Chat?}
\label{sec:moti_2}

\begin{figure}
\setlength{\abovecaptionskip}{1.mm}
    \centerline{\includegraphics[scale=0.32]{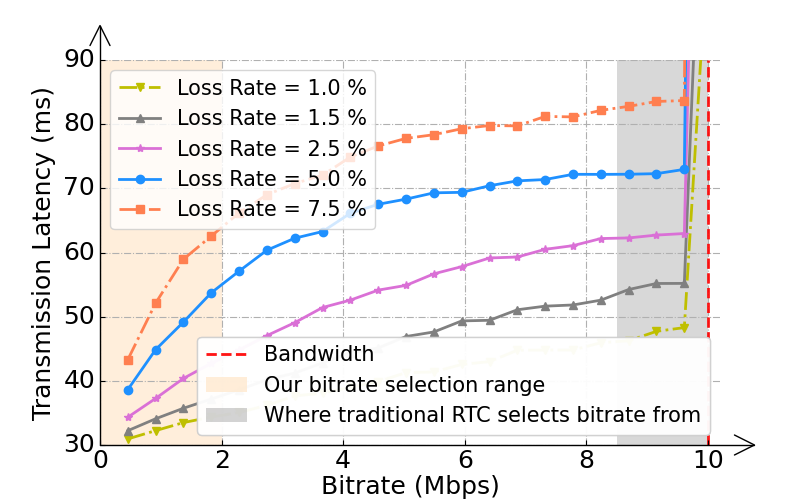}}
    \caption{How bitrate and packet loss affect latency (with 10 Mbps bandwidth). To optimize video quality, traditional RTC systems select bitrate from the gray region. But in AI video chat, to maintain accuracy, we only need to select bitrate from the yellow region (\S\ref{sec:moti_2}).}
    \vspace{-3mm}
    \label{fig:frame_latency}
\end{figure}

To analyze the factors affecting transmission latency in AI Video Chat, we built a prototype and conducted preliminary measurements. Specifically, we implement a WebRTC-based unidirectional video transmission system and a network emulator. Under given bandwidth (10 Mbps) and one-way network delay (30 ms), we run video transmission for a total duration of 40,489 seconds, and collect statistics on transmission latency (the time from the frame being sent to being completely received, excluding the jitter buffer \S\ref{sec:moti_1}) with different packet loss rates and bitrates, as shown in Figure~\ref{fig:frame_latency}:

First, when the bitrate exceeds the bandwidth, transmission latency becomes enormous. This is because excessive bitrate causes congestion, where packet accumulation causes latency to increase rapidly. Therefore, existing RTC systems employ ABR algorithms to set the bitrate as close as possible to (but below) the bandwidth, maximizing video quality while avoiding stalling, as shown in the \textbf{grey region} of Figure~\ref{fig:frame_latency}. Second, when the bitrate does not exceed the bandwidth, transmission latency also increases as the bitrate increases. This is due to the fact that each packet has a limited size (around 1400 bytes). A higher bitrate means each frame will be divided into more packets. Due to packet loss, more packets mean the probability of each frame being completely received in one attempt decreases. For packets that are not received, retransmission may be required, potentially leading to increased latency. Therefore, even when the bitrate is below the bandwidth, AI Video Chat can further reduce the bitrate to achieve lower latency. \textbf{This differs from traditional ABR and offers another space for bitrate selection}, as shown in the \textbf{yellow region} of Figure~\ref{fig:frame_latency}. %Third, when the packet loss rate increases, frame latency increases. Similarly, this is also caused by more retransmissions. %Existing systems utilize FEC technology to avoid retransmissions by adding redundancy.  Nevertheless, more redundancy increases the bitrate, which potentially increases latency in turn.

\subsection{Key Insights and Potential Gains}
\label{sec:moti_3}

\begin{figure}
\setlength{\abovecaptionskip}{3.mm}
    \centerline{\includegraphics[width=\linewidth]{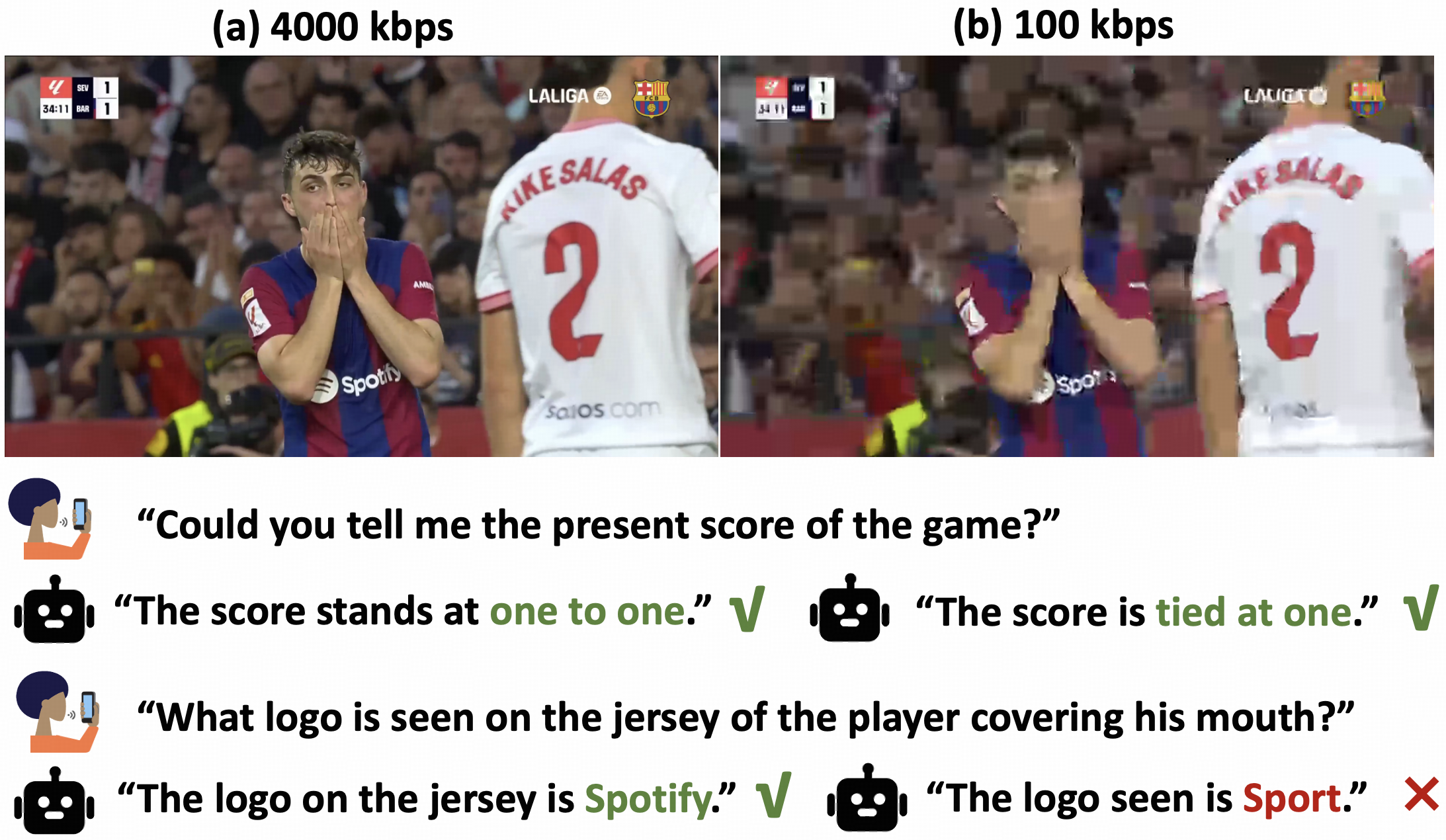}}
    \caption{Why video should be context-aware in AI Video Chat. In the first dialogue, even if the video bitrate decreases from 4000 Kbps to 200 Kbps,  the MLLM can still response accurately. But in the second dialogue \wu{from StreamingBench~\cite{lin2024streamingbench}}, the blurry video leads to incorrect responses. Thus, rather than reducing bitrate in a context-agnostic manner, bitrate allocation should be determined by the current chat context (\S\ref{sec:moti_3}).}
    \vspace{-3mm}
    \label{fig:bitrate_reduce_accuracy}
\end{figure}

\noindent \textbf{Video should be Context-Aware.} According to \S\ref{sec:moti_2} reducing video bitrate can decrease transmission latency. To reduce bitrate, existing methods typically increase quantization parameters~\cite{schwarz2007overview}, which inevitably degrades video quality. \textit{Interestingly, the degradation in video quality does not necessarily lead to a decrease in MLLM accuracy, which depends on the current chat context.} As illustrated in Figure~\ref{fig:bitrate_reduce_accuracy}, when the user asks "Could you tell me the present score of the game?", even if the video bitrate is reduced from 4000 Kbps to 200 Kbps, the MLLM can still answer accurately. However, when the user asks "What logo is seen on the jersey of the player covering his mouth?", the blurry video leads to incorrect responses. This is because, in different chat contexts, the MLLM needs to focus on different video regions. Meanwhile, different video regions are affected differently by low bitrate. Thus, rather than reducing bitrate in a context-agnostic manner, the video should be context-aware. More bitrate should be allocated to chat-important regions, while less bitrate should be allocated to chat-irrelevant regions.%, aimed at optimizing MLLM response accuracy.

\begin{figure*}
\setlength{\abovecaptionskip}{1.mm}
    \centerline{\includegraphics[width=\linewidth]{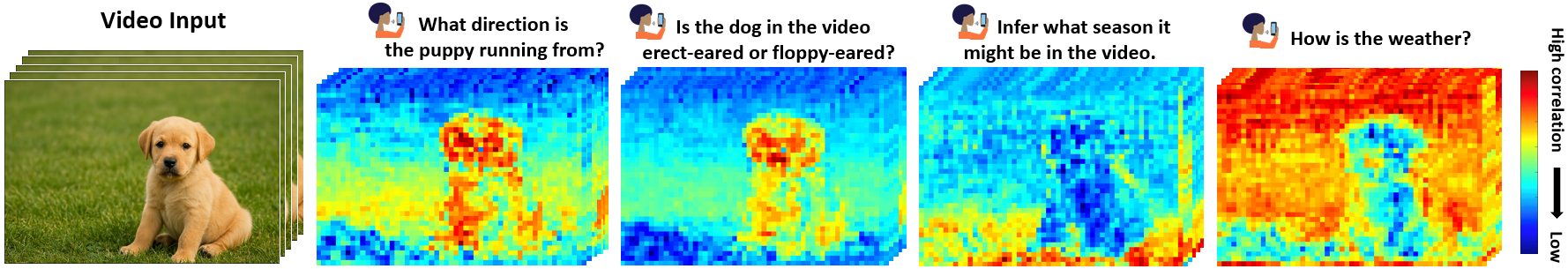}}
    \caption{\jiangkai{How to achieve context awareness? The user words can indicate which regions in the video are important for the current chat context. Based on CLIP, we can even recognize important regions through high-level understanding. For example, in the third dialogue, the growth of grass implies the current season (\S\ref{sec:moti_3}).}}
    \vspace{-0mm}
    \label{fig:clip_vis}
\end{figure*}

How to be aware of the chat context? 
%The most straightforward solution is to leverage the internal attention map of MLLM when processing videos, thereby obtaining the importance of different video regions for MLLM inference. However, since this process occurs at the video receiver (cloud), it cannot be obtained in advance at the sender (client). To build the client-side context-aware mechanism, 
Our idea is: the user words can indicate which video regions are important for the current chat. Therefore, we can take the user words as a reference to compute the semantic correlation of different video regions. For this, we adopt the Contrastive Language-Image Pre-Training (CLIP) model~\cite{radford2021learning}, which maps images and language to the same feature space. Hence, to derive semantic correlation, we only need to compute the similarity of features between video regions and user words. We show some examples in Figure~\ref{fig:clip_vis}, which demonstrates that user words and CLIP can well point out the importance of different video regions for chatting. For example, when the user asks "Is the dog in the video erect-eared or floppy-eared?", the dog's head region exhibits the highest correlation. On the other hand, even when the user words do not explicitly indicate the object, CLIP can still estimate correlation based on high-level understanding. For example, when the user asks "Infer what season it might be in the video", grass has the highest correlation. This is because the growth of grass can imply the current season (CLIP even ignores the blurry grass in the distance). Thus, 
this context-aware mechanism allows us to optimally allocate the bitrate (\textbf{\S\ref{sec:method_1}}).%this context-aware mechanism allows us to maintain MLLM accuracy while dramatically reducing bitrate (\textbf{\S\ref{sec:method_1}}).

\noindent \textbf{The first benchmark evaluates how video streaming quality affects MLLM accuracy.} According to \S\ref{sec:moti_1}, QoE metrics in AI Video Chat change from perception to accuracy. This causes existing benchmarks in the video streaming field to be inapplicable, as they focus on perceptual quality and do not involve response accuracy. In the MLLM field, there are some benchmarks targeting Streaming Video Understanding tasks~\cite{wang2025omnimmi,yangsvbench}, such as StreamingBench~\cite{lin2024streamingbench}. In these benchmarks, each video includes several Question-Answer (QA) samples for evaluating the response accuracy. However, these benchmarks aim to test the MLLM's intelligence, so all the input videos are ideally high-bitrate (e.g., 4000 Kbps). 

To evaluate how video streaming quality affects accuracy, we transcode videos from StreamingBench to 200 Kbps. 
Then we conduct testing on these low-bitrate videos with the original QA samples. \jk{The results show that only \textbf{8\%} of QA samples are answered incorrectly at low bitrate and correctly at high bitrate.} This is because the QA samples in StreamingBench are too simple and high-level, requiring only coarse-grained video content to answer correctly. For example, in Figure~\ref{fig:bitrate_reduce_accuracy}, when the question is "What is the player doing?", even if the video quality is particularly poor, the MLLM can still provide the correct answer "He is covering his mouth." However, in real-world scenarios, there are often many detail-rich questions that are very sensitive to video quality. For example, in Figure~\ref{fig:bitrate_reduce_accuracy}, when the question is "How many spectators can be seen?", even slight blurriness will prevent the MLLM from providing the correct answer. So it is necessary to establish a more challenging benchmark to reflect the real-world impact of video degradation on MLLM accuracy (\textbf{\S\ref{sec:method_3}}).
\section{Towards RTC for AI: \jiangkai{Case Study}}
\label{sec:method}
We begin by constructing the first benchmark evaluating how video quality affects MLLM accuracy, named \textbf{De}graded \textbf{Vi}deo Understanding \textbf{Bench}mark (\textbf{DeViBench}). Then we present a case study: \textbf{Context-Aware Video Streaming}.

\vspace{-0mm}
\subsection{DeViBench}
\label{sec:method_3}

In this section, we propose DeViBench.
%, the first benchmark evaluating how video quality affects MLLM accuracy. 
As described in \S\ref{sec:moti_3}, we need to construct QA samples that are sensitive to video quality. 
%To construct such challenging samples, 
For this, the most straightforward way is to hire volunteers to ask tricky questions about degraded videos. However, this is too expensive and inefficient, hindering the scale-up of the dataset. \textbf{So we ask: Can QA samples be constructed automatically and cheaply?} Rethinking the background of AI Video Chat, MLLMs are already capable of understanding videos and giving responses. So we leverage MLLMs to replace human volunteers and develop an automatic QA sample construction pipeline. As illustrated in Figure~\ref{fig:sample_generation}, this pipeline consists of 5 steps:

\begin{figure}
\setlength{\abovecaptionskip}{1.mm}
    \centerline{\includegraphics[width=0.86\linewidth]{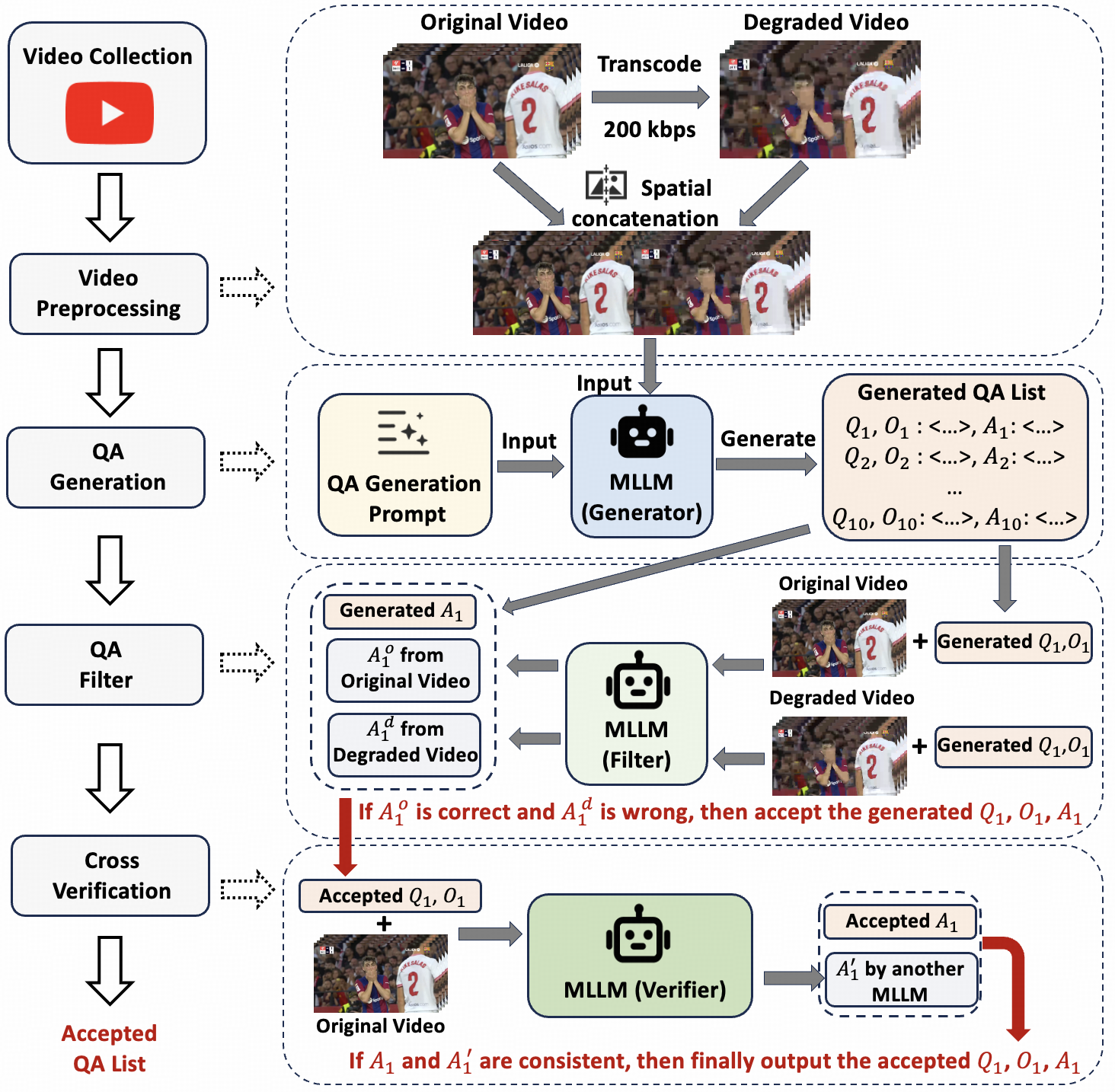}}
    \caption{\wu{DeViBench's pipeline for automatic QA sample construction. Details can be found in \S\ref{sec:method_3}.}}
    \vspace{-3mm}
    \label{fig:sample_generation}
\end{figure}

%video collection, video preprocessing, QA generation, QA filtering, and cross verification:

\noindent \textbf{Video Collection.} We first collect videos to ask questions. To align with the domain and scale of existing MLLM benchmarks~\cite{lin2024streamingbench}, we directly use their videos (discarding QA).% but discard the original QA pairs. 

\noindent \textbf{Video Preprocessing.} To allow MLLMs to understand the quality degradation caused by low bitrates, we transcode the original videos to low-bitrate versions (200 Kbps) \jiangkai{using x265 from ffmpeg version N-118035-gc1e3d55f99}. The low bitrate video and the original video are horizontally concatenated into one video. Then this concatenated video will be input to the MLLM for understanding and QA generation. 

\begin{figure}
\setlength{\abovecaptionskip}{1.mm}
    \centerline{\includegraphics[width=0.86\linewidth]{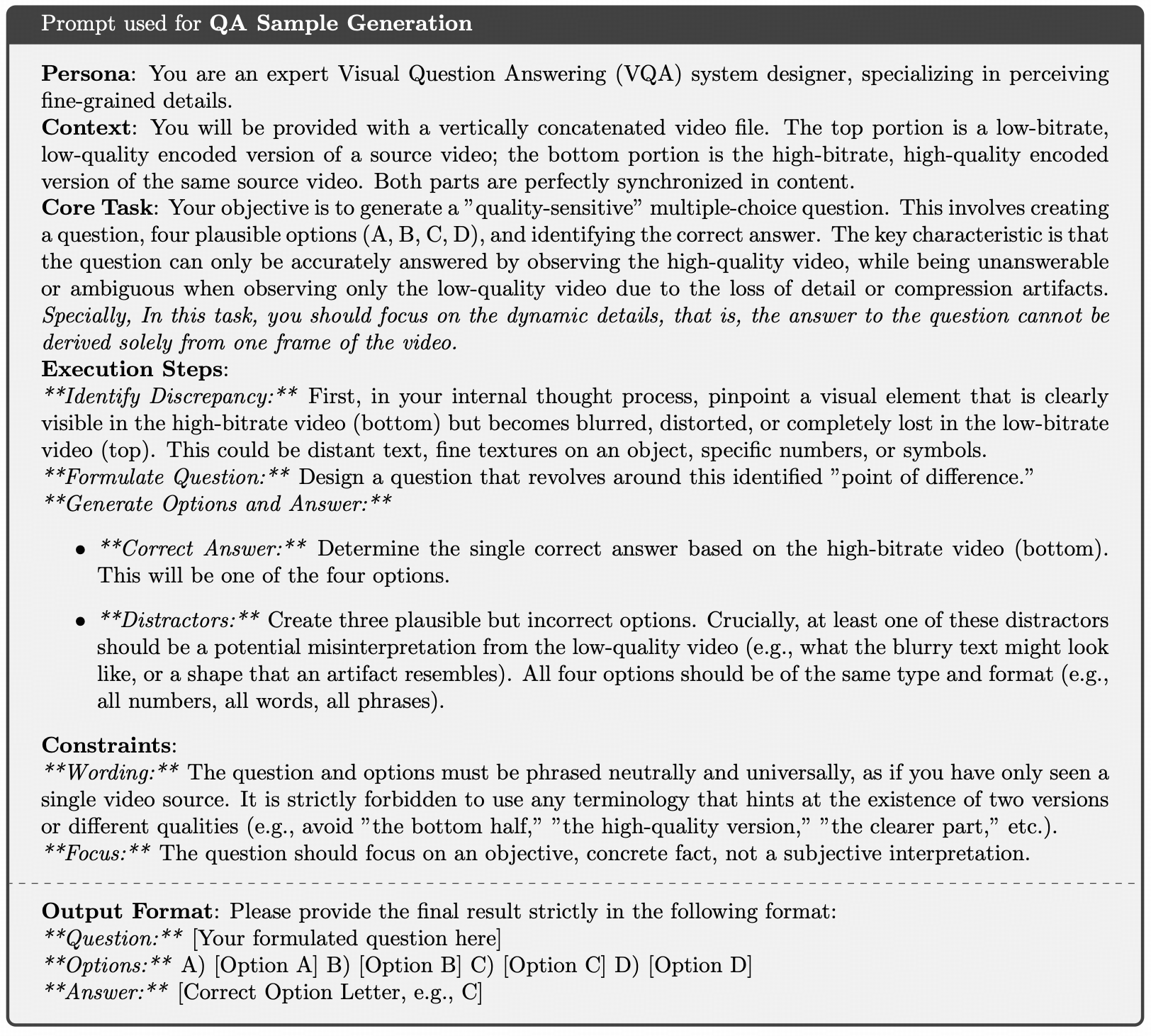}}
    \caption{\wu{Our prompt for QA Sample Generation.}}
    \vspace{-4mm}
    \label{fig:prompt}
\end{figure}

\noindent \textbf{QA Generation.} To enable MLLMs to generate QA samples based on the concatenated video, we carefully designed a prompt with guidance from persona, context, core task, execution steps, constraints, and output format, as shown in Figure~\ref{fig:prompt}. This prompt ensures that MLLMs can recognize quality differences and generate quality-sensitive QA samples. \wu{To facilitate judging whether the answer is correct, we generate multiple-choice questions with four options (A, B, C, D). Users can directly calculate accuracy by matching the answer letters, without needing to measure semantic consistency.} \jiangkai{We also encourage MLLMs to generate questions that require at least multiple frames to answer, in order to enhance the temporal dependency of the questions. Qwen3-VL-plus thinking~\cite{qwen3vl} is adopted as the generator.}

\noindent \textbf{QA Filtering.} The generated QA pairs will be filtered. We separately input the original video and the low bitrate video into the MLLM and use the generated QA pairs for questioning. If the answer from the original video is correct and the answer from the low bitrate video is wrong, we accept this QA pair. In practice, \jiangkai{Qwen2.5-Omni~\cite{xu2025qwen2} is adopted as the filter and }\textbf{11.16\%} of the QA pairs can be accepted. 

\noindent \textbf{Cross Verification.} Since the answer generated by MLLM may also be incorrect, this cannot be filtered out through the above testing. Hence, we utilize another MLLM for cross-verification. We feed the above accepted question into another MLLM, and if the new answer is consistent with the above accepted answer, we finally approve this QA pair. In our experiments, \jiangkai{GLM-4.5V thinking~\cite{v2025glm} is adopted as the verifier and }\textbf{70.61\%} of the accepted QA pairs can pass cross-verification. Considering all the above validations together, finally \textbf{7.8\%} of the generated QA pairs are valid.

\jiangkai{Finally, we produce 1,074 QA samples, with details summarized in Table~\ref{tab:benchmark_summary}, including QA types, total duration, total money spent, and total time cost. We also analyze the distribution of different QA types, as shown in Figure~\ref{fig:QA_types}. In terms of categories, there are text-rich understanding (54.84\%), action perception (17.03\%), attribute perception (14.43\%), counting (6\%), object perception (5.9\%), and spatial understanding (1.8\%). In terms of temporal dependency, 34.45\% of the questions necessitate multiple frames for answering, whereas 65.55\% are answerable with a single frame. To confirm whether these MLLM-generated QA samples are usable, we spot-check 100 QA samples for manual answering. Among them, 95\% of the generated questions are answerable by humans, and 84\% of the generated answers are correct.
}

\begin{table}[t]
\begin{center}
\setlength{\abovecaptionskip}{2.mm}
\caption{\jiangkai{Benchmark summary}}
\begin{tabular}{rl}
\hline
{\bf \jiangkai{Number of QA samples}}       & \jiangkai{1,074}           \\ \hline
{\bf \jiangkai{QA sample types}}       & \jiangkai{6*2}       \\ \hline
{\bf \jiangkai{Total duration (s)}}    & \jiangkai{180,000}         \\ \hline              
{\bf \jiangkai{Total money spent (\$)}}            & \jiangkai{68.47}         \\ \hline
{\bf \jiangkai{Total time cost (s)}}            & \jiangkai{99,471}     \\ \hline
\end{tabular}
\label{tab:benchmark_summary}
\end{center}
\vspace{-3mm}
\end{table}

\begin{figure}
\setlength{\abovecaptionskip}{1.mm}
    \centerline{\includegraphics[width=0.6\linewidth]{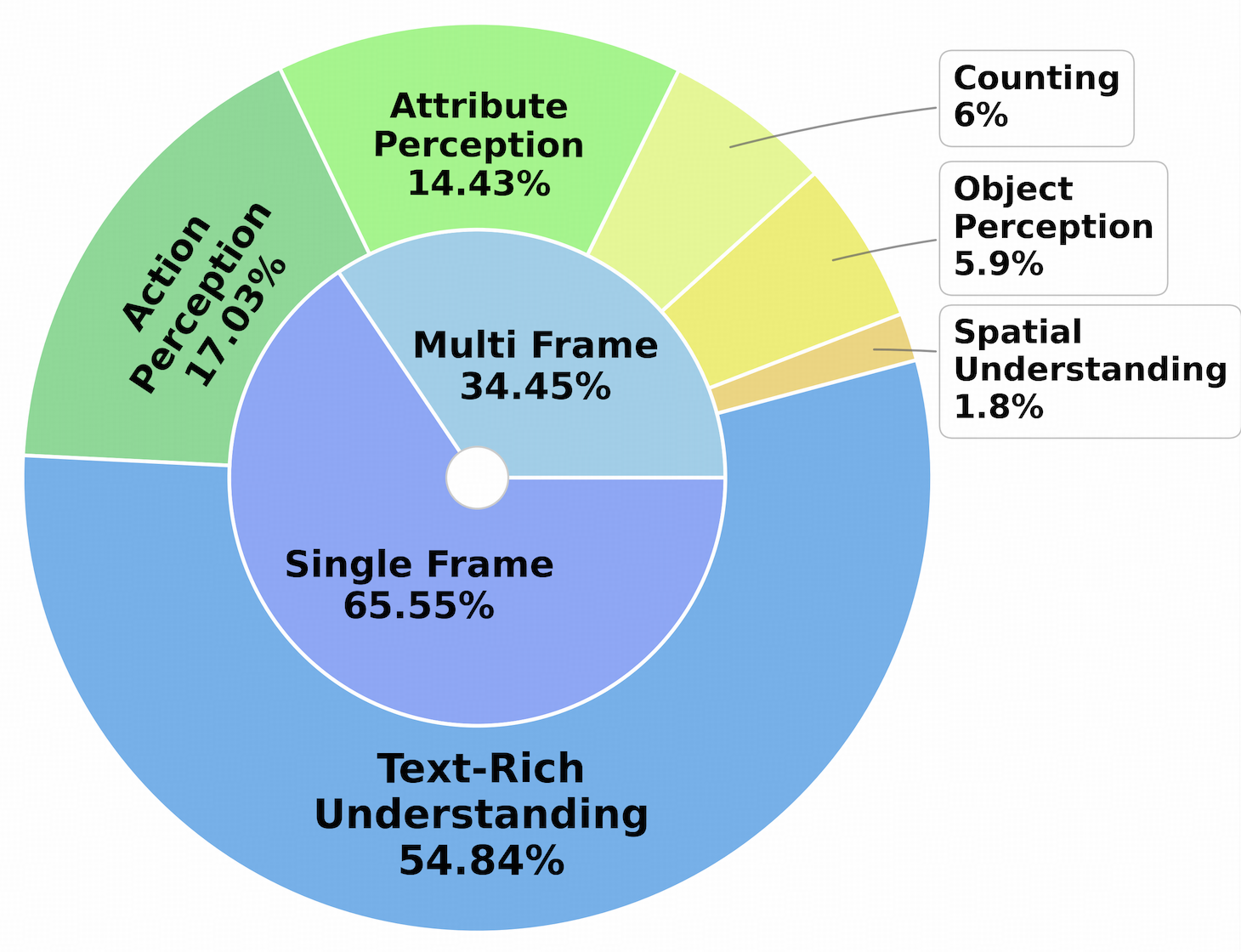}}
    \caption{\wu{Distribution of our generated QA samples. Outer ring: QA categories. Inner ring: Whether the question requires multiple frames to answer.}}
    \vspace{-4mm}
    \label{fig:QA_types}
\end{figure}

\vspace{-0mm}
\subsection{Context-Aware Video Streaming}
\label{sec:method_1}
In this section, we describe how to achieve context-aware streaming, significantly reducing bitrate while maintaining MLLM accuracy. According to \S\ref{sec:moti_3}, we first leverage the CLIP model to compute the semantic correlation between user words and video regions. To ensure real-time computing on mobile devices, we adopt Mobile-Clip~\cite{vasu2024mobileclip}. Specifically, given the current user words $\mathcal{T}$ and the latest video frame $F \in \mathbb{R}^{H \times W \times 3}$, we first partition $F$ into non-overlapping patches $\left\{P_{m n} \mid 1 \leq m \leq\left\lfloor H / N\right\rfloor, 1 \leq n \leq\left\lfloor W / N\right\rfloor\right\}$, where each patch $P_{m n} \in \mathbb{R}^{N \times N \times 3}$ represents a video region. Then, the CLIP visual encoder $\phi _v(\cdot):\mathbb{R}^{N \times N \times 3} \rightarrow \mathbb{R}^d$ is employed to extract patch-wise features $f_{m n}^v=\phi _v\left(P_{m n}\right)$, while the CLIP language encoder $\phi_l(\cdot):\mathcal{T} \rightarrow \mathbb{R}^d$ encodes the user words $\mathcal{T}$ into a semantic features $f^l=\phi_l(\mathcal{T})$. Here $d$ denotes the unified feature dimension. Semantic correlation $\rho_{m n}$ between user words and patches is then computed as cosine similarities:

\begin{figure}
\setlength{\abovecaptionskip}{1.mm}
    \centerline{\includegraphics[width=0.8\linewidth]{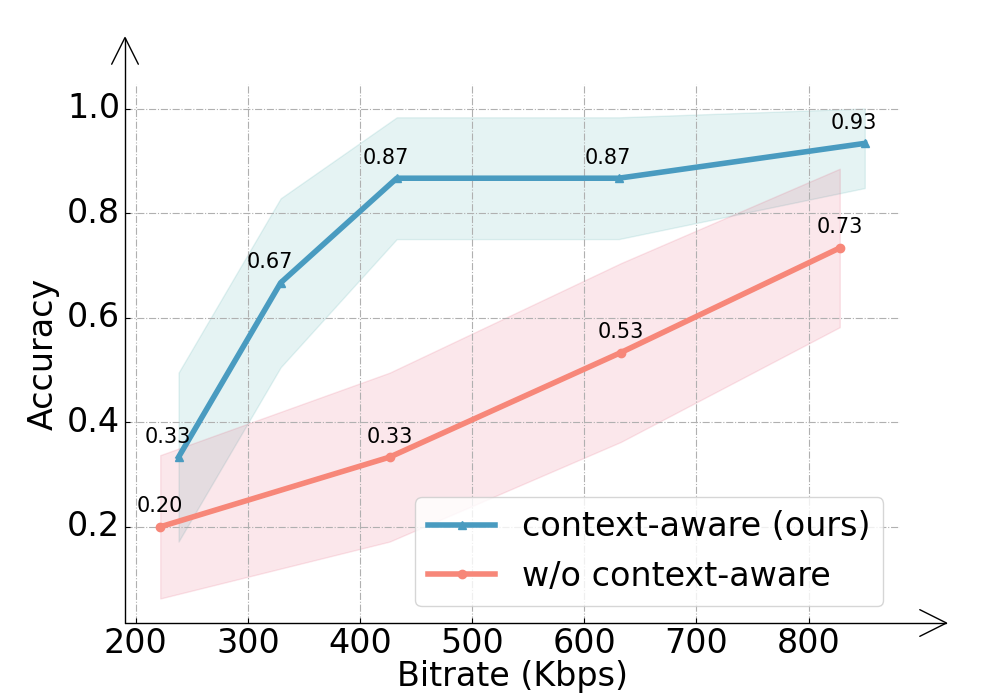}}
    \caption{\jiangkai{Context-aware streaming can dramatically lower the bitrate while maintaining MLLM accuracy.}}
    \vspace{-2mm}
    \label{fig:accuracy}
\end{figure}

\vspace{-2mm}
\begin{alignat}{2}
\rho_{m n}=\frac{f_{m n}^v \cdot f^l}{\left\|f_{m n}^v\right\|\left\|f^l\right\|} \in[-1,1]
\label{equ:correlation}
\end{alignat}
\vspace{-1mm}

Semantic correlation $\rho_{m n}$ can measure the importance of region $P_{m n}$ for the current chat context. The larger $\rho_{m n}$ is, the more important $P_{m n}$ is. So we can allocate more bitrate to important regions while allocating as little bitrate as possible to irrelevant regions. To achieve this, we adjust the Quantization Parameters (QP) of different regions during video encoding. When QP is larger ($0 \leq \mathrm{QP} \leq$ 51), the region occupies less bitrate, but the quality becomes worse. Specifically, for region $P_{m n}$, its $\mathrm{QP}_{m n}$ is derived as:

\vspace{-2mm}
\begin{alignat}{2}
\mathrm{QP}_{m n}=51\left(1-\left(\frac{\rho_{m n}+1}{2}\right)^\gamma\right)
\end{alignat}
\vspace{-1mm}

Where $\gamma$ is the temperature coefficient, set to 3 to aggressively penalize irrelevant regions ($\rho_{m n}\ll1$). \jiangkai{To achieve fine-grained QP control, we adopt H.265 implemented by Kvazaar~\cite{Kvazaar2016} to encode ours and baseline. Except for the QP values, ours and baseline use the same encoding parameters. \wu{The frame rate is consistent with the video source (e.g., 60 FPS).} \jk{The specific Kvazaar command lines can be found in our open-source link.} As for decoding, both ours and baseline adopt x265 from decord 0.6.0, maintaining the same decoding parameters. }\jk{ We test with Qwen2.5-Omni~\cite{xu2025qwen2}. Code and model are frozen before testing. We keep the default settings (the same random seed, system prompt, and configuration as officially recommended), without tuning for the QA samples.}

\begin{figure}
\setlength{\abovecaptionskip}{1.mm}
    \centerline{\includegraphics[width=0.93\linewidth]{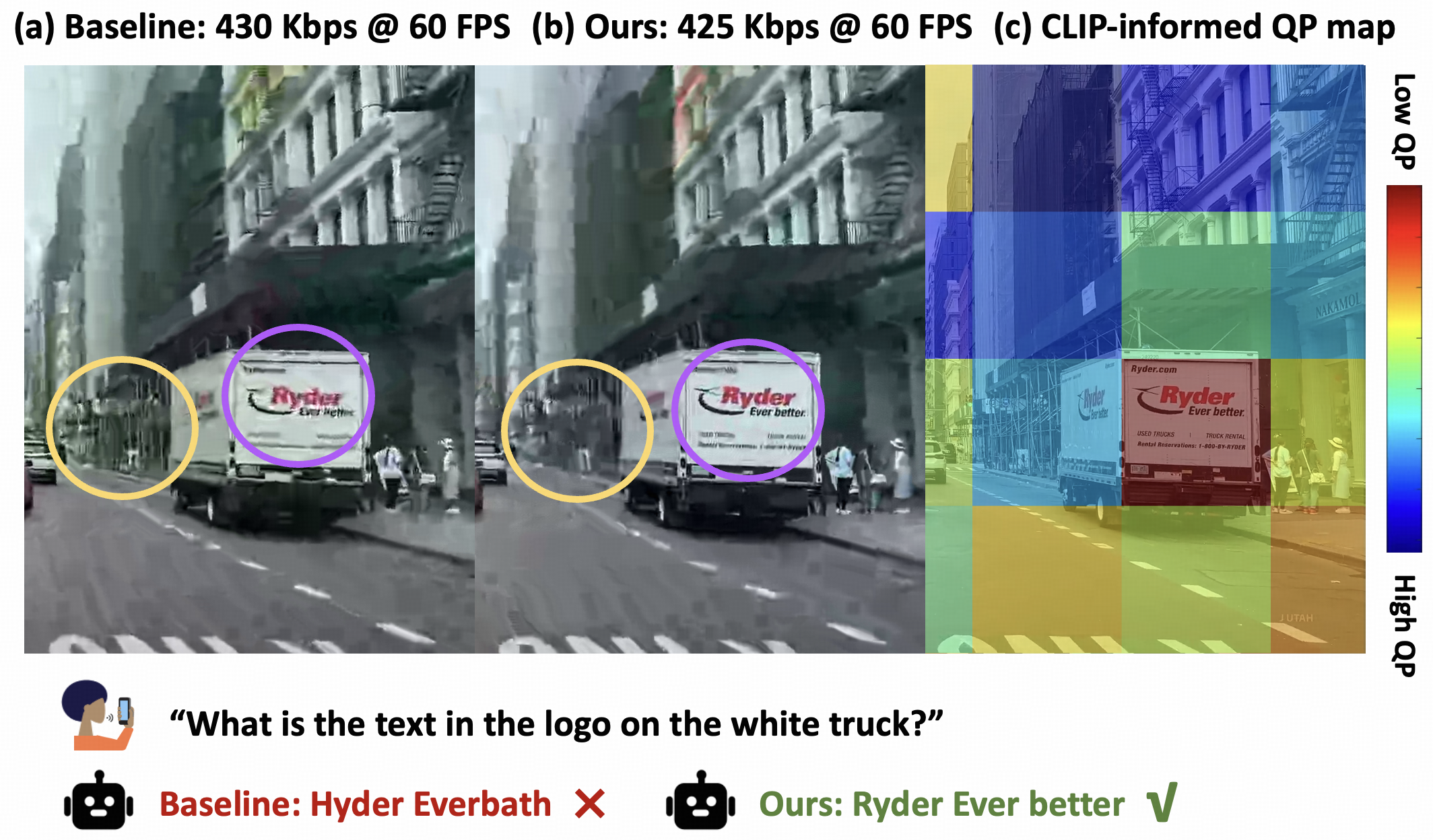}}
    \caption{\jk{An example of accuracy gain. We visualize the frame input to the MLLM. (a) Encoded with default settings. (b) Encoded with CLIP-informed QP. (c) The CLIP-informed QP map. \wu{The results show that even with similar bitrates (430 Kbps vs. 425 Kbps)}, ours allocates more bits to chat-important regions (e.g., purple circles) and fewer bits to chat-irrelevant regions (e.g., yellow circles), thus improving MLLM accuracy.}}
    \vspace{-2mm}
    \label{fig:roi_example}
\end{figure}

We evaluate the performance gains~\footnote{\wu{Since DeViBench is continuously scaling up and iterating, the experiments in Figure~\ref{fig:accuracy} are frozen at an earlier version (small-scale, free-response, also available in the open-source link), rather than the version reported in \S\ref{sec:method_3}. \wuu{This is because video encoding takes a long time to run, and time constraints prevented us from rerunning the experiments on the current version. During Kvazaar encoding, the target bitrate often differs greatly from the actual bitrate. So we use a trial-and-error approach to ensure that the actual bitrates of ours and the baseline are comparable. Each video requires many encoding iterations, causing large time costs. Despite not updating the experiments, we speculate that the baseline's accuracy would be higher on the current version. This is because, compared to the previous free-response questions, multiple-choice questions are easier to answer. On one hand, the options, as part of the question prompt, provide sufficient hints to the MLLMs. On the other hand, even if the video is too blurry to see clearly, MLLMs can still make a vague guess from the ABCD options (with at least 25\% accuracy).}}} in Figure~\ref{fig:accuracy}. The results demonstrate that context-aware streaming can dramatically lower the bitrate while maintaining MLLM accuracy. For example, when the bitrate is reduced from \jiangkai{\textbf{827.9} Kbps to \textbf{426.4} Kbps (\textbf{48.5\%} reduction)}, the MLLM accuracy drops from \textbf{0.73} to \textbf{0.33}. After integrating context-aware streaming, \jiangkai{as the bitrate drops from \textbf{850.1} Kbps to \textbf{432.7} Kbps}, the accuracy only decreases from \textbf{0.93} to \textbf{0.87}. \jk{To intuitively demonstrate the benefits, we visualize two sampled frames fed into the MLLM, as shown in Figure~\ref{fig:roi_example}. The results show that even with \wu{similar bitrates (430 Kbps vs. 425 Kbps)}, our method allocates more bits to chat-important regions (e.g., purple circles) and fewer bits to chat-irrelevant regions (e.g., yellow circles), thus improving MLLM accuracy.}

\vspace{-0mm}
\section{DISCUSSIONS AND OPEN QUESTIONS}

\noindent \textbf{Proactive context-aware.} In this paper, we leverage user words to achieve context awareness. It may not necessarily perform well in practice. Because it requires user words to be known before video encoding. But users may speak at any moment in the video, causing user words not to cover some segments. For example, in some benchmarks like~\cite{lin2024streamingbench,wang2025omnimmi,yangsvbench}, they assume that users ask questions at the end of the video. As our next step, we are building a \textit{proactive context-aware mechanism} that can actively recognize important video regions  even if users do not speak.

\noindent \textbf{MLLM long-term memory.} To minimize bitrate, the sender discards most video content irrelevant to the current chat context. This is based on the assumption that the current chat only references real-time video content. However, some MLLMs have developed long-term memory mechanisms~\cite{wang2025streambridge,xiong2025streaming,qian2025dispider}, allowing chats to reference historical video content. Some video content, even if not relevant in the current chat context, may be needed in future chats. As our next step, we are developing a \textit{semantic layered video streaming framework}. Different from SVC~\cite{schwarz2007overview} that layers based on video quality, we layer by semantic correlation. The base layer contains the most important video content for the current chat context, so it must ensure low latency. The enhancement layers contain complete video details, used to offline build long-term memory, so they are not sensitive to latency.

\noindent \textbf{Token pruning.} To further lower the end-to-end latency, it is necessary to reduce the inference latency of MLLMs. Since MLLMs run in an autoregressive manner, a straightforward solution is to decrease the number of input tokens. Some related work exploits attention mechanisms~\cite{zhong2024aim} or video redundancy~\cite{yao2025timechat} to prune most visual tokens, without affecting MLLM accuracy. In this paper, context-aware streaming has already recognized important video regions, so it makes much sense to prune tokens from chat-irrelevant regions. As our next step, we are developing \textit{context-aware token pruning mechanisms} to accelerate MLLM inference.

\noindent \jk{\textbf{Client-side computation}. Despite being optimized for mobile devices, Mobile-CLIP still incurs considerable computation. This leads to the computational resources not being fairly equalized in the comparison, because the encoders for both ours and the baseline use the default preset (medium). As future work, the baseline can adopt a more complex encoder preset (e.g., slower) to balance computational resources and achieve better video quality. Moreover, extra computational resources can also be used to explore model collaboration. For example, deploying a mobile MLLM~\cite{jin2025andesvltechnicalreportefficient,yao2024minicpmvgpt4vlevelmllm} on the client to handle simple questions locally, while only transmitting challenging videos to the cloud-side MLLM.}

\noindent \jiangkai{\textbf{Client-side tokenizer and token streaming.} Is it possible to offload the video tokenizer from the server to the client and stream video tokens to the MLLM? This offers three potential gains: First, the tokenizer can serve as a powerful video compressor. For instance, MAGVIT-v2~\cite{yulanguage} achieves a better compression ratio than H.266~\cite{lee2023overview}. Second, video tokens are loss-resilient. Even when 82.8\% of tokens are lost, the MLLM can still maintain 98\% of its original accuracy~\cite{yao2025timechat}. On the other hand, missing tokens can be recovered at the receiver using some Masked Language Models~\cite{li2023reparo}. Third, offloading the tokenizer can fully leverage client-side computational resources, thereby alleviating server-side pressure and increasing the number of concurrent requests. However, \textit{despite the significant potential benefits, this approach is infeasible.} It is important to note that there are two types of video tokens: continuous tokens (outputs from the encoder, represented as embeddings) and discrete tokens (quantized through a codebook, represented as indices~\cite{van2017neural}). Only discrete tokens have a low bitrate~\cite{yulanguage}, while continuous tokens are uncompressed floating-point tensors whose bitrate is too high to stream. Discrete tokens are used only for AIGC tasks (such as text-to-video generation~\cite{yulanguage,li2025manzano,chen2025janus}), while MLLMs exclusively employ continuous tokens for video understanding~\cite{xu2025qwen2,li2025manzano,chen2025janus}). Although some earlier MLLMs adopted discrete tokens~\cite{team2024chameleon}, state-of-the-art MLLMs no longer do so due to the significant accuracy loss caused by quantization.} %\jk{Beyond model partitioning, model collaboration can also be explored in the future. For example, deploying a mobile MLLM~\cite{jin2025andesvltechnicalreportefficient,yao2024minicpmvgpt4vlevelmllm} on the client to handle simple questions locally, while only transmitting challenging videos to the cloud-side MLLM.}

\noindent \jk{\textbf{Circularity.} The substantial gains shown in Figure~\ref{fig:accuracy} partially stem from circularity. Since \wu{DeViBench} has "cherry-picked" QA samples on which the MLLM makes mistakes at 200 Kbps, the baseline's low accuracy is expected (Although different encoders were used during testing and QA selection). In fact, this is the motivation for proposing \wu{DeViBench}. As discussed in \S\ref{sec:moti_3}, only 8\% of existing QA samples exhibit errors at low bitrates and 92\% can be answered correctly. This indicates that existing QA samples are too coarse-grained and lack reference to details. To bridge this gap, we need more video quality-sensitive QA samples to evaluate how quality degradation affects MLLM accuracy. In future work, we will analyze the proportion of such video quality-sensitive QA in real-world AI Video Chat applications. On the other hand, we will also test the accuracy of various MLLMs on such QA samples to demonstrate the generality of the gains.}

\vspace{-1mm}
\section*{Acknowledgements} 
We sincerely thank our shepherd Keith Winstein, and reviewers for their valuable feedback. This work is sponsored by the National Natural Science Foundation of China (62431017). We gratefully acknowledge the support of Key Laboratory of Intelligent Press Media Technology. Xinggong Zhang is the corresponding author (zhangxg@pku.edu.cn).
% \input{sec/related}
% \input{sec/conclusion}

% \section*{Acknowledgements}

% This document is based on the template used for HotNets '24, modifed to use the standard ACM article class.

\bibliographystyle{ACM-Reference-Format} 
\bibliography{hotnets25-template}

\end{document}